\documentclass[sigconf]{acmart}
\usepackage{algorithm}
\usepackage{algpseudocode}
\usepackage{bbm}
\usepackage{multirow}
\usepackage{makecell}

\algrenewcommand\algorithmicrequire{\textbf{Input:}}
\algrenewcommand\algorithmicensure{\textbf{Output:}}
\algnewcommand\algorithmicforeach{\textbf{for each}}
\algdef{S}[FOR]{ForEach}[1]{\algorithmicforeach\ #1\ \algorithmicdo}

\AtBeginDocument{%
  \providecommand\BibTeX{{%
    \normalfont B\kern-0.5em{\scshape i\kern-0.25em b}\kern-0.8em\TeX}}}

\begin{document}

\title{Towards Efficient Pareto-optimal Utility-Fairness between Groups in Repeated Rankings}

\author{Phuong Dinh Mai}
\email{phuong.dm2@vinuni.edu.vn}
\affiliation{%
  \institution{College of Engineering \& Computer Science, \\ VinUniversity}
  \city{Hanoi}
  \country{Vietnam}
}

\author{Duc-Trong Le}
\email{trong.ld@vnu.edu.vn}
\affiliation{%
  \institution{University of Engineering and Technology, \\ Vietnam National University}
  \city{Hanoi}
  \country{Vietnam}}
  
\author{Tuan-Anh Hoang}
\email{anha2k47@gmail.com}
\affiliation{%
  \institution{School of Science, Engineering \& Technology \\ RMIT University Vietnam}
  \city{Hanoi}
  \country{Vietnam}}

\author{Dung D. Le}
\authornote{Corresponding author}
\email{dung.ld@vinuni.edu.vn}
\affiliation{%
  \institution{College of Engineering \& Computer Science, \\ VinUniversity}
  \city{Hanoi}
  \country{Vietnam}
}

\renewcommand{\shortauthors}{Trovato and Tobin, et al.}
\renewcommand\footnotetextcopyrightpermission[1]{}
\settopmatter{printacmref=false} 

\begin{abstract}
In this paper, we tackle the problem of computing a sequence of rankings with the guarantee of the Pareto-optimal balance between (1) maximizing the utility of the consumers and (2) minimizing unfairness between producers of the items. Such a multi-objective optimization problem is typically solved using a combination of a scalarization method and linear programming on bi-stochastic matrices, representing the distribution of possible rankings of items. However, the above-mentioned approach relies on Birkhoff-von Neumann (BvN) decomposition, of which the computational complexity is $\mathcal{O}(n^5)$ with $n$ being the number of items, making it impractical for large-scale systems. To address this drawback, we introduce a novel approach to the above problem by using the Expohedron — a permutahedron whose points represent all achievable exposures of items. On the Expohedron, we profile the Pareto curve which captures the trade-off between group fairness and user utility by identifying a finite number of Pareto optimal solutions. We further propose an efficient method by relaxing our optimization problem on the Expohedron's circumscribed $n$-sphere, which significantly improve the running time. Moreover, the approximate Pareto curve is asymptotically close to the real Pareto optimal curve as the number of substantial solutions increases. Our methods are applicable with different ranking merits that are non-decreasing functions of item relevance. The effectiveness of our methods are validated through experiments on both synthetic and real-world datasets.
\end{abstract}

\maketitle

\section{Introduction}
In the ever-expanding digital landscape, recommender systems play an essential role in shaping user experiences, influencing choices, and connecting individuals with numberless products and content. As these systems continue to evolve and proliferate, there is a growing recognition of the ethical considerations surrounding their design and impact. One of such considerations is \emph{producer fairness}, which arises from the desire to create trustworthy recommendation experiences for users while fostering a certain level playing field for service providers or content producers.

{\bf Challenge.} Striking an optimal balance between the producer fairness and the user utility becomes a nuanced challenge, where the system aims not only to enhance the relevance and satisfaction of the users, but also to minimize disparities and ensure unbiased treatment across diverse item groups.

In addressing this complex landscape, some studies advocate for constraint handling as a strategic approach, introducing specific fairness criteria to guide and control the optimization process. Sinan Seymen \cite{SeymenAM21a} has proposed a comprehensive framework based on a bi-stochastic probability matrix, offering a structured methodology for tackling the intricate interplay between user utility and fairness considerations. Alternative methodologies, such as reinforcement learning, have gained prominent attention \cite{Ge_2021}, \cite{Ge_2022}, \cite{Popcorn}, among others. Despite the diverse array of approaches, profiling the Pareto front \cite{ParetoIntro} emerges as a preferred strategy for decision makers. Representing a set of non-dominated solutions, where enhancing one objective comes at the expense of another, the Pareto front provides decision-makers with the ability to explore and extract solutions that epitomize the optimal compromise between user utility and fairness. This is important for decision-makers navigating the intricate landscape of recommender system optimization, offering a practical and effective means to address the inherent trade-offs between user-centric utility and the imperative of fairness.

Scalarization techniques, such as weighted sum of objective functions, are commonly used to obtain the Pareto front. They enable the exploration of different trade-offs by adjusting the scalarization parameters. However, in repeated ranking, these techniques usually depend on a bi-stochastic probability matrix followed by an expensive Birkhoff-von Neumann (BvN) \cite{Dufosse2016} decomposition to get a distribution of item orders. This decomposition algorithm does not scale well \cite{Geyik_2019} when it involves a graph matching algorithm. Using state-of-the-art linear programming solvers adapted to our problem, the total complexity of such an approach is then $O(n^5)$.

{\bf Approach.} Inspired by the Expohedon concept introduced in \cite{Kletti_2022}, our solution contributes to the ongoing exploration of geometric solutions for the fairness-utility problem in recommendation systems. In this work, we profile the entire Pareto-optimal solution spaces in the context of the producer fairness—user satisfaction problem on the Expohedron facets. We leverage this exploration with the concept of geodesic paths within a hypersphere to approximate this frontier. While previous endeavors have explored the use of geometry for this purpose, to the best of our knowledge, we are pioneering an approach to approximate the exact Pareto set using the geodesic of a hypersphere \cite{Gotoh94}. This innovative perspective aims to advance the field's understanding and computational efficiency in resolving the intricate trade-offs between producer fairness and user satisfaction within recommender systems. This approximation provides a notable advantage by significantly reducing the computational time required and employing Carathéodory decomposition in \cite{Kletti_2022}, which is more efficient and resource-conserving compared to the BvN approach, for the distribution of orders. An additional benefit of this approximation is the ability to control the trade-off between the accuracy of the approximation and the overall running time which will be essential for limited resources scenarios. 

Our main technical contributions are:
\begin{enumerate}
    \item We profile the Pareto-optimal solutions space for the producer fairness and user utility in repeated rankings' problem on the Expohedron polytope. We consolidate our theory by perform an experiment in a small synthetic dataset. 
    \item With the aforementioned profiling, we introduce an approximate solution based on the geodesic of a hypersphere. This approach not only offers a time-efficient alternative but also enables us to exchange algorithmic accuracy for enhanced runtime efficiency. 
    \item We perform experiments in two real-world dataset: TREC Fair Ranking dataset \cite{biega2021overview} and MSLR \cite{DBLP:journals/corr/QinL13} to compare our methods to quadratic programming and learning to rank baselines in terms of Pareto-optimality and efficiency.
\end{enumerate}

\section{Related Work}
In this section, we review closely related work and provide some detailed background that will be used throughout the paper.

\textit{Fairness in ranking}. Fairness in recommender systems is a crucial consideration aimed at addressing biases and ensuring equitable treatment among users and content producers. As these systems shape user choices, there's a growing emphasis on mitigating algorithmic discrimination and promoting transparency. Methods for addressing biases in recommender systems are categorized into three types: pre-processing, in-processing, and post-processing. Pre-processing methods, exemplified by \cite{salimi2019capuchin} and \cite{10.5555/3322706.3362016}, focus on manipulating data before it enters the ranking model to remove sensitive information and mitigate biases. In-processing methods, such as \cite{Wang2020DCNVI}, involve modifying the ranking within the learning-to-rank framework. Post-processing methods, like those described in \cite{Zerveas2022MitigatingBI}, are popular for their flexibility and adaptability, involving re-ranking strategies after the initial recommendations. The presented work follows a post-processing approach in addressing bias in recommender systems.

Fairness considerations in recommender systems can be approached from various perspectives, including the standpoint of consumers, as discussed in \cite{Do_Corbett-Davies_Atif_Usunier_2022}, or from the viewpoint of item producers, as explored in \cite{Kletti_2022} and \cite{Wang2020DCNVI}. The focus of our work lies specifically in the context of item producers, offering insights and solutions that address fairness concerns from the producers' standpoint.

We adhere to the premise that the item positioned at the top of a list garners more attention from consumers. Consequently, popular items consistently occupying the top positions may perpetuate the 'Rich Gets Richer' phenomenon, leaving a long tail of less-attended or newly-released items behind \cite{Naghiaei2022TheUO}. This trend is commonly observed in large e-commerce recommender systems (RS). In response, there is a growing interest in addressing the fairness of exposure. Rather than ensuring all items in the system receive a uniform level of exposure \cite{Biega_2018}, our objective is to ensure that all groups of items receive a predetermined level of exposure—focusing on fairness at the group level. This aligns with the approach taken by Singh and Joachims \cite{Singh_2018}. 

\textit{Position Based Model (PBM)}. In fairness of exposure approach, re-ranking models depend on the amount of exposure each item received at a certain ranking position. With PBM \cite{book}, the amount of exposure only depends on the ranking itself. Other models might have the ranking depending on items at higher ranks (i.e. Dynamic Bayesian Network \cite{book}). Our work is applicable to PBMs.

\textit{Pareto optimal}. In the realm of recommendation algorithms, Pareto's optimality \cite{ParetoIntro} holds significance within the business market in general and the recommender systems in particular, serving as a cornerstone in the pursuit of optimal trade-offs between competing objectives. This concept is pivotal as it allows recommender systems to navigate the intricate balance between conflicting objectives such as user satisfaction and system efficiency. Furthermore, Pareto optimality facilitates the exploration of complex decision spaces, enabling recommender systems to adapt dynamically to evolving user preferences and system constraints. The integration of Pareto optimal solutions empowers recommender systems to deliver personalized and equitable recommendations that enhance user experience and drive overall system performance. However, we believe that this field have not received enough attention from other researchers. Recent studies surrounding this trade-off curve in recommendation include \cite{Jin2023ParetobasedMR}, \cite{10.1007/978-981-16-2594-7_60}, \cite{Paparella2022PursuingOT}, etc. 

\textit{Doubly bi-stochastic matrix}. Singh and Joachims \cite{Singh_2018} formulate the ranking distribution as a doubly bi-stochastic matrix whose elements represent the probabilities of each items' appearance in each ranking position. In PBMs, user's utility could be formulated as a linear objective, such as Discounted Cumulative Gain (DCG) \cite{DCG} or the Rank-biased precision \cite{RPB}. The fairness could be formulated as a linear constraint, so the problem is solved using Linear Programming (LP) \cite{LP}. The probability matrix is then decomposed by Birkhoff–von Neumann (BvN) decomposition \cite{Dufosse2016} to form an expected value of a distribution over permutations. The same approach could be seen in later research such as \cite{su2021optimizing}, \cite{Wang_2021}. The limitation of this approach is it could only generate one single point in the Pareto optimal set—the point at the fairness target level. 

Nevertheless, using a scalarization technique, fairness could be formulated into a second quadratic/linear objective combining with user's utility function by trade-off hyperparameter $\alpha$. By changing the hyperparameter $\alpha \in [0, 1]$, we are able to generate a full set of Pareto solution set. This method is compatible with our proposed method and will be used as a baseline. 

\textit{Controller}. This is a heuristic approach using a greedy algorithm. Whenever a query is repeated, an error terms is added to the relevance scores, which will be arranged in the next ranking. It starts with the PBM, reduces the empirical unfairness at every time-step by giving adequate bonuses to disadvantaged items and after $T$ repeated time, the expected exposure of all groups might be equal. The static setting \cite{MRFR} will also be used as a baseline for our work.  

\textit{Expohedron}. Our work is strongly relied on in the notion of Expohedron in \cite{Kletti_2022}. Based on the definition of the Permutahedron \cite{Permutahedron}, an Expohedron is a convex polytope whose vertices represent all attainable exposure vector with one ranking in PBMs. With this, our feasible solution region reduces its dimension from $\mathbb{R}^{n \times n}$ in the Birkhoff polytope to $\mathbb{R}^{n}$ with $n$ is the number of items need to re-rank. A point inside this polytope represents an expected exposure vector of distribution over all ranking. Kletti et al. \cite{Kletti_2022} also proposed Carathéodory decomposition — an algorithm allows an Expohedron internal point to be factorized into a distribution over vertices, which implies permutations.  

\section{Problem and Preliminary}
\subsection{Problem and formulation}
\textit{Settings.} In our analysis, each query is treated as an independent entity. Consider a specific query, denoted as $q$, which corresponds to a collection of $n$ items indexed by $i$. A specialized black-box search and scoring mechanism provides us with a column vector $\rho$, residing within the interval $[0, 1]^n$. This vector elucidates the extent of relevance each item possesses in relation to the query. Notably, each item is categorized into one specific group. Given the potential repetition of each query, we seek to discern the anticipated trade-off curve between user satisfaction and the disparity between the item categories. It is pertinent to mention that the binary matrix indicator $G \in \mathbbm{R}^{n \times g}$ denotes whether the item $i$ is affiliated with a group $j$, with the constraint $\sum_i G_{ij} = 1$ ensuring exclusive membership to a single group.

\textit{Ranking matrix.} Let $S_n \in \mathbbm{R}^{n \times n}$ be a matrix representing permutations, where $||S_n|| = n! = N $ denotes the total number of permutation matrices. Within this set, let $\pi \in S_n$ represent a specific permutation matrix. The entry, $\pi_{ij} = 1$ if and only if the item $i$ occupies rank $j$. Let $\mathbbm{1}$ be a column vector of 1, the bi-stochastic matrix $\pi$ is formulated as:
\begin{equation*}
\begin{split}
    \pi &\in \{0, 1\}^{n \times n} \\
    st &\; \mathbbm{1}_n^T \pi = \mathbbm{1}_n^T\\
    &\; \pi \mathbbm{1}_n = \mathbbm{1}_n
\end{split}
\end{equation*}

\textit{PBMs.} In our empirical analyses, we employ the Position Based Model \cite{book}, represented by a column vector $\gamma \in \mathbbm{R}^n_+$ whereas the $k^{th}$ value denote the exposure associated with the ranking $k$. Without compromising the generality of our analysis, we posit that the vector 
$\gamma$ exhibits a decreasing trend as the rank increases.

With this model, the item' exposure score at ranking $\pi$ is represented by $E(\pi) = \pi \gamma$ whereas the $k^{th}$ element of $E(\pi)$ denotes the exposure value of the item $k$.

Since the query is repeated, assume the distribution $\mathbbm{D}$ of $N$ rankings $\pi_1, \pi_2, \cdots, \pi_N$ in $S_n$ is delivered with proportion $p_1, p_2, \cdots, p_N$. The expected exposure score vector associated with items is:
\begin{equation} \label{ori_exposure}
    \begin{split}
        E(\mathcal{D}) &= \sum_i^N p_i E(\pi_i) = \sum_i^N p_i \pi_i \gamma \\
        st & \sum_i^N p_i = 1; \; 0 <= p_i= < 1
    \end{split}
\end{equation}

\textit{Utility.} The user' satisfaction score or the utility of a ranking $\pi$ is denoted as $U(\pi) =  E(\pi) \rho = \pi \gamma \rho$. Relatively, the expected utility function of the distribution $\mathbbm{D}$ is:
\begin{equation}
\begin{split}
    U(\mathcal{D}) &=  \sum_i^N p_i U(\pi) = \sum_i^N p_i \pi \gamma \rho \\
    &= E(\mathcal{D}) \rho
\end{split}
\end{equation}

When $\gamma_k = \frac{1}{\log_2{k+1}}$ for position $k$ in the list, $U(\pi)$ denotes the Discounted Cumulative Gain \cite{RPB}. Alternatively, when $\gamma_k = (1-r)r^{k-1}$ with $r \in (0, 1)$, $U(\pi)$ becomes the Rank-biased precision \cite{RPB} metric. 

\textit{Unfairness.} In addressing the multifaceted nature of fairness, we operate under the premise that a decision-maker has delineated what constitutes fair exposures. Specifically, a target exposure is established for each group, drawing inspiration from works such as \cite{Diaz_2020}, wherein the target exposure is conceptualized as an affine function of the relevance vector $\rho$, embodying a form of meritocratic fairness. This target quantifies the exposure a group is deemed deserving every time the query is invoked. Meeting this predetermined exposure threshold signifies system fairness by definition. Conversely, to quantify the degree of unfairness within a system, we  necessitate a metric that gauges the deviation from this target exposure. In this paper we proceed as in \cite{Diaz_2020} in adopting the standard Euclidean norm: 
\begin{equation} 
    F(\mathcal{D}, \epsilon^*) = ||G E(\mathcal{D}) - \epsilon^*||_2
\end{equation}

This measurement allows us to trade off utility with the fairness.

\textit{\bf Objectives.} Both $U$ and $F$ are contingent on $\mathcal{D}$ solely through its expectation. Thus, we can express them as $U(\mathcal{D})$ and $F(\mathcal{D}, \epsilon^*)$ instead of $U(E(\mathcal{D}))$ and $F(E(\mathcal{D}), \epsilon^*)$ for short. Our objective is to address a Multi-Objective Optimization (MOO) challenge with dual goals: maximizing utility and simultaneously minimizing unfairness, particularly after processing an extensive array of rankings.
\begin{equation} \label{objs}
    \max_{\mathcal{D}} \; U(\mathcal{D}), \; \min_{\mathcal{D}} \; F(\mathcal{D}, \epsilon^*)
\end{equation}

\subsection{Expohedron}
\textit{The permutation simplex}. A naive way to solve problem (\ref{objs}) is to calculate the distribution vector $p$ representing $\mathcal{D}$. In this method, our variable is a vector $p \in \mathbbm{R} ^ {n!}$, which is infeasible in reality due to the high dimensionality of the space $\mathbbm{R} ^ {n!}$. 

\textit{Birkhoff polytope}. In the PBMs, we can work in a smaller, more tractable space: the space of bi-stochastic matrices as known as the Birkhoff polytope \cite{Singh_2018}. The expected exposure in \ref{ori_exposure} could be represented as:
\begin{equation}
    \begin{split}
        E(\mathcal{D}) &= \sum_i^N p_i \pi_i \gamma \; = \; (\sum_i^N p_i \pi_i) \gamma \\
        &= B\gamma \\
        st &\; \mathbbm{1}_n^T B = \mathbbm{1}_n^T\\
    &\; B \mathbbm{1}_n = \mathbbm{1}_n
    \end{split}
\end{equation}
The bi-stochastic property of the matrix $B$ could be easily proved by calculating the sum of each column/row so we defer the proof of this property to interested readers. Our problem is rewritten into:
\begin{equation} \label{BvN}
\begin{split}
    \max_B & \; U(\mathfrak{D}) = B \gamma \rho \\
    \min_B & \; F(\mathfrak{D}) = ||G B \gamma- \epsilon^*||_2 \\
    st &\; \mathbbm{1}_n^T B = \mathbbm{1}_n^T\\
    &\; B \mathbbm{1}_n = \mathbbm{1}_n
\end{split}
\end{equation} 

The Birkhoff polytope is synonymous with the convex hull delineated by all permutations, each of which is represented by a permutation matrix. A notable algorithm, termed the Birkhoff-von Neumann (BvN) decomposition \cite{Dufosse2016}, facilitates the representation of any bi-stochastic matrix of dimensions $n \times n$ as a convex combination comprising a maximum of $(n-1)^2+1$ permutation matrices. The complexity of the BvN is $O(max(n^{d+2}, n^4)) \approx O(n^5)$ with $O(n^d)$ is the complexity of the algorithm for finding a matching in graphs on $n$ vertices (number of items), plus $O(n^2)$ to manipulate the doubly bi-stochastic matrix.

\textit{Expohedron}. Geir Dahl \cite{MajorizationTheory} showed that the feasible region of $E(\mathcal{D})$ is a majorization permutation \cite{majorizarion}. Let the expected exposure $E(\mathcal{D})$ is the vector variable $x$, our problem is formulated as:
\begin{equation} \label{expo_prob}
    \begin{split}
        \max_x & \; U(x) = x \rho \\
        \min_x & \; F(x) = ||G x - \epsilon^*||_2 \\
        st & \; x \preceq \gamma
    \end{split}
\end{equation}
This polytope is also known as the Expohedron  which Till Kletti et al. \cite{Kletti_2022} used to solve the individual fairness in repeated ranking problem. In the settings of PBMs, the Expohedron inherits multiple properties of the permutahedron \cite{Permutahedron} which we will discuss further when solving our group fairness for repeated ranking problem.

Moreover, Till Kletti et al.\cite{Kletti_2022} also showed that the decomposition algorithm using Carathéodory’s theorem \cite{Carathodory1907berDV} can express any point $x$ inside
the Expohedron over at most $n$ rankings in the $O(n^2 \log n)$ running time.

\section{Profiling the Pareto-optimal solutions space}
\subsection{Theoretical analysis}
In this section, we show how to find the exact solution for the multi-objectives problem (\ref{expo_prob}) in section 3. We will start will the most important notion of our proof.

\hfill\\
\textbf{Theorem 1}: The Pareto optimal solutions of the problem (\ref{expo_prob}) lie on Expohedron facets. These facets have a dimension of at most $(n-g-1)$.

\underline{\textit{Proof}}:

Considerate the level of fairness $F(x) = ||\epsilon||_2 = f_0$, the Pareto solution associated with this level of fairness is the solution of this optimal problem:
\begin{equation} \label{optimal_at_fairness}
    \begin{split} 
        \max_x & \; U(x) = x \rho \\
        st & \; G x - \epsilon^* = \epsilon \\
        st & \; x \preceq \gamma
    \end{split}
\end{equation}
Assume that problem (\ref{optimal_at_fairness}) has at least one optimal solution. The fact that the majorization condition “$\preceq$” could be written in the form sum of largest elements in an array makes this condition is completely linear. Overall, both conditions and the optimal function are linear. Therefore, the problem (\ref{optimal_at_fairness}) is a typical linear programming (LP) problem. 

The theorem 3.3 in \cite{LP} states that if the objective function has a maximum/minimum value on the feasible region, then it has this value on (at least) one of the extreme points. The feasible region of 
(\ref{optimal_at_fairness}) is an intersection of a sub-plane with a convex polytope A which is also a convex polytope B. Consequently, There is an optimal solution to be the vertex of B and laid on the facet of A. 

Since $G$ is a binary matrix, $R^{g} = Gx - \epsilon^*$ must be parallel with at least 2 facets of the Expohedron (facets of the Expohedron could be written in form of binary matrix multiplication equals to sum of some largest elements in an array). Therefore, the vertex where $R^{g}$ intersecting with the Expohedron boundary lie on the facet with the dimension of at most $n-g-1$. 

Before delving into the optimal Pareto solution, we will establish several lemmas utilizing our self-defined geometric objects.

\hfill\\
\textbf{Lemma 1}: The $L_2$ norm fairness level in group-fairness setting are “hyper-cylinders” formulated by $S^{n-g-1} \times R^{g}$ ($g$ is the number of groups). 

\underline{\textit{Proof.}}: The feasible space for the linear problem $C = \{x \; \mid \; Gx = \epsilon^*\}$ is the subspace $\mathbbm{G}$ dimension $g$ of the $\mathbbm{R}^n$. Assume $\widebar{\mathbbm{G}}$ is the orthogonal subspace of $\mathbbm{G}$ dimension $n-g$ of the $\mathbbm{R}^n$. Projecting all $\mathbbm{G}$ onto $\widebar{\mathbbm{G}}$ creates a point $\bar{C}$ in $\widebar{\mathbbm{G}}$. 

Furthermore, the set of points which have equidistant in $\mathit{L_2}$ from $\bar{C}$ form a hypersphere in $\widebar{\mathbbm{G}}$ or so-called the $(n-g-1)$-sphere $S^{n-g-1}$ \cite{topology}. 

Therefore, the $L_2$ norm fairness level in group-fairness setting are “hyper-cylinders” formulated by $S^{n-g-1} \times R^{g}$ ($g$ is the number of groups).

\hfill\\
\textbf{Lemma 2}: An intersection of the "hyper-cylinder" $H = S^{n-g-1} \times R^{g}$ with a subplane $\alpha$ in $\mathbbm{R}^n$ returns:
\begin{enumerate}
    \item A "hyper-cylinder" $\widehat{H} = S^{n-g-2} \times R^{g}$ if $\alpha$ is a hyperplane containing the sub-plane $R^{g} = \{x \; \mid \; Gx = \epsilon^*\}$
    \item A "hyper-ellipsoid" if a sub-plane $\alpha$ dimension $(n-g)$ intersects with the sub-plane $C = \{x \; \mid \; Gx = \epsilon^*\}$
\end{enumerate}

\underline{\textit{Proof}}: 
\begin{enumerate}
    \item Since $\alpha$ is a hyperplane containing $R^{g}$, $\alpha$ is orthogonal with  $\widebar{\mathbbm{G}}$, we could write $\alpha = R^{n-g-1} \times R^{g}$. It is easy to prove that the intersection of a d-dimensional sphere and a hyperplane is a d-1 dimensional sphere using the Pythagorean theorem. Therefore, the intersection of $\alpha$ with $H$ could be written in the form of:
    \begin{equation}
    \begin{split}
        I &= (S^{n-g-1} \cap R^{n-g-1}) \times R^{g} \\
        &= S^{n-g-2} \times R^{g} \\
    \end{split}
    \end{equation}
    \item Assume $P_0$ is a $(n-g-1)$-sphere representing the intersection of $H$ with $\widebar{\mathbbm{G}}$. Since $\alpha$ has the same dimension with $\widebar{\mathbbm{G}}$ and intersect $C$, we could rewrite $\alpha = \{\widebar{\mathbbm{G}}A$ with A is a transformation full rank matrix.
    Therefore, the intersection of $H$ and $\alpha$ could be written as:
    \begin{equation}
        \begin{split}
            P &= P_0 A \\
            &= S^{n-g-1} A
        \end{split}
    \end{equation}
    The image of an n-sphere under an invertible affine transformation defines a hyper-ellipsoid. This theorem could be proved using the Spherical coordinates \cite{ARFKEN200383}. 
\end{enumerate}

Figure \ref{prob with n=3, g=2} is an example of the optimal Pareto solutions when n=3 and g=2.
\begin{figure}
    \centering
    \includegraphics[scale=0.5]{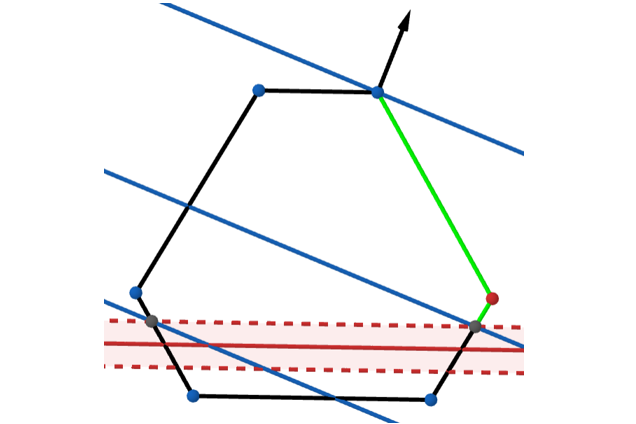}
    \caption{An example of utility and fairness level set when n=3 and g=2. While the blue line is the utility level set, the target fairness level is the red ones and one set of fairness level are red dashed line. The green line segments form the Pareto-curve in the Expohedron}
    \label{prob with n=3, g=2}
\end{figure}

\hfill\\
\textbf{Theorem 2}: Let $P_1$ and $P_2$ be an optimal Pareto solution at facet $F$ having dimension $(n-g)$. The segment between $P_0$ and $P_1$ is part of the Pareto optimal solution. 

\underline{\textit{Proof}}: 
Using all aforementioned Lemma, let point $P_0$ be the intersection of the hyperplane $P = \{\textbf{x} \mid \sum_i^n x_i = \sum_j^n \gamma_j\}$ and the target fairness level $U_0$ and the facet $F$. The unfairness level $U_1$ containing $P_1$ intersects with the hyperplane $P$ at the hyper-cylinder $H_1$ and this hyper-cylinder $H_1$ intersects with the facet $F$ having dimension $(n-g)$ at the hyper-ellipsoid $E_1$. The same notion with $P_2$. 

Considerate to hyper-ellipsoid $E_1$ containing $P_1$. If $P_1$ is a Pareto-optimal points, $P_1$ must have the largest utility value amongst $E$. This means the tangent vector at $P_1$ of $E$ must be parallel with the relevance vector $\rho$ - which is equal to the orthogonal to the hyperplanes of utility level. Furthermore, $E_1$ and $E_2$ have similar center as well as affine transformation matrix (in the same cutting facet $F$). Therefore, $E_2=s E_2$ with $s$ is a scaling scalar. So without the loss of generality, the vector $\vec{P_0 P_1} = s \vec {P_0 P_2}$ defines the optimal utility-unfairness trade-off solution direction in $F$.

Moreover, the \textbf{Theorem 1} shows that, without others facets restriction, the optimal solution $P_i$ always lies on $F$. It's apparent that segment $P_1P_2$ is a part of optimal Pareto solution.

With the establishment of \textbf{Theorem 1} and the \textbf{Theorem 2}, its inevitable consequence is the following corollary.  
\hfill \\ \\
\textbf{Corollary}: The optimal Pareto solution for our problem is a set of consecutive segments across Expohedron facets. These segments connected by a "break" point laid on the boundary of the facet. At these points, our optimal direction changes the direction to go to another facets.  \label{corollary}

\begin{figure}
    \centering
    \includegraphics[scale=0.4]{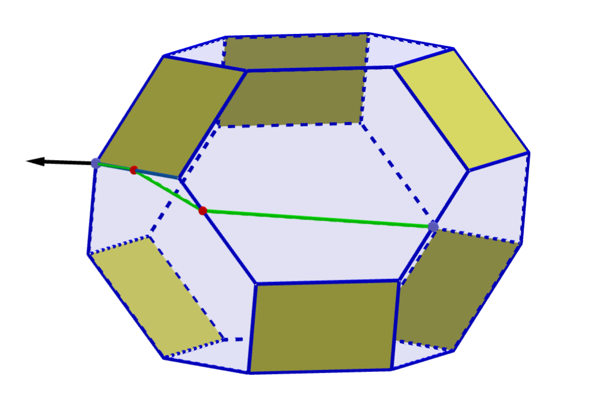}
    \caption{An example of utility and fairness level set when n=4. The green line segments form the Pareto-curve in the Expohedron. The red points are “break” points in Pareto set curve}
    \label{prob with n=4}
\end{figure}

\subsection{Finding the Pareto curve}
The analysis presented above provides a straightforward solution: Begin at the point of optimal utility on the target fairness level and move along the optimal direction of $\theta$ on that facet until intersecting the boundary of the Expohedron. This intersection denotes a 'break' point. The subsequent facet to enter will offer the optimal trade-off between utility and unfairness for each unit increase. By iteratively repeating this process, we ensure reaching a point that maximizes utility, serving as the endpoint of the Pareto curve.

\begin{algorithm} \label{alg_exact}
\caption{The P-Expo for profiling the Pareto-optimal solutions space on the Expohedron}
\begin{algorithmic}[1]
\Require 
    \Statex $\rho \in (0, 1)$: vector of relevance 
    \Statex $G \in \mathbb{R}^{n \times n\_group}, \: G_{ij} \in \{0, 1\}$: item masking group 
    \Statex $\varepsilon \in \mathbb{R}^{n\_group}$: the target group fairness 
    \Statex $\gamma$: the position based exposure
\Ensure
    \Statex Pareto Set - $P$
    \Statex

\State $G_k \gets \sum_{i=1}^k \gamma_{i}^{\uparrow}$
\State $opt_F = point \gets \text{Solve (\ref{optimal_at_fairness}) for } \epsilon^*$
\State $P \gets \{opt_F\}$
\While {True}
    \State $s \gets \sum_{i=1}^k point_{i}^{\uparrow}$
    \State $S \gets \{k_1, k_2, ...\} \gets$ which $G_k == s_k$ 
    \State $face\_orth \gets $ binary matrix from invert permutation of $S_k$ 
    \State $slope \gets 0$
    \State $\theta \gets (0, 0, ..., 0)$
    \ForEach {$vec \in face\_orth$ or $vec = \emptyset$} 
        \State $adj\_face \gets \{face\_orth \setminus vec\}$
        \State $\theta_f \; slope_{f} \gets $ Find the optimal direction and its trade-off in $adj\_face$ 
        \If {$slope < slope_{f}$}
            \State $slope = slope_p$
            \State $\theta = \theta_f$
        \EndIf
    \EndFor
    \State $next\_point \gets  \text{Expohedron.intersect(}point, \theta\text{)}$
    \State $P$.append($next\_point$)
    \State $point \gets next\_point$
\EndWhile
\end{algorithmic}
\end{algorithm}

To determine the value of $\theta$ at the facet $F$, let's assume the existence of an optimal point $P_1$. According to \textbf{Theorem 2}, we can identify another optimal point $P$ situated on $F$ or at the intersection of the target fairness level and the facet $F$. We then define $\theta$ as the vector $\vec{PP_1}$.

\section{Approximation of the Pareto curve} 
The PBMs' Expohedron has a very appealing feature. It is an inscribing polytope, a polytope in which all its vertex lies on an n-sphere. Considering the hyperplane dimension $n-1$ $\sum_i^n x_1i =\sum_i^n \gamma$, it is easy to see that all vertices are equidistant from the center point coordinator $(c, c,...,c) \in \mathbb{R}^n: \;c = \frac{\sum \gamma}{n}$). 

\begin{figure}
    \centering
    \includegraphics[scale=0.5]{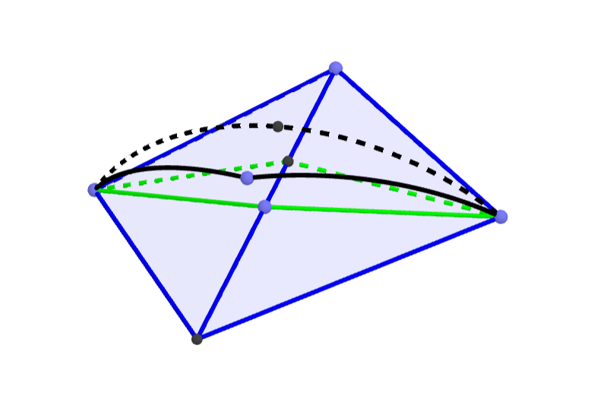}
    \caption{An example of how a set of connected segment could be approximate with set of geodesics. The dash is the approximate geodesic curve and its projection on Expohedron, while the line is the exact solution.}
    \label{geodesic approximation}
\end{figure}

Snyder's research \cite{Snyder2012MapPA} showed that geodesic on hypersphere transforming to straight lines via the central projection, so-called gnomonic projection in 2-dimension. With the use of the corollary in section \ref{corollary}, the Pareto-optimal solutions space could be mapped to a set of consecutive geodesic on the $(n-2)$-sphere. Although the 2 consecutive geodesics $\widearc{AB}$ and $\widearc{BC}$ could not be summed up to make the geodesic $\widearc{AC}$, all segments in the Pareto-optimal solutions set are linear transformations from the item relevance vector. Therefore, we assume that the geodesic $\widearc{AC}$ could be used to approximate the 2 consecutive geodesics $\widearc{AB}$ and $\widearc{BC}$. Though the tangent vector of this approximation is not exactly matched with those two, it still shows the same trend with the optimal solutions in the trade-off curve. Figure (\ref{geodesic approximation}) is an example. Since the function for the hypersphere geodesic is proven in \cite{bishop2013riemannian}, the cost for the sampling process on the geodesic is $\mathcal{O}(1)$.

In order to make the approximation more accurate, we could use some marked points to split the geodesic curve into multiple ones. And when the number of marked points is large enough, each projection of the geodesic will be a segment in one hyperplane and contains no “break” point, which means it is part of the Pareto set. A marked point could be found by solving a QP problem:
\begin{equation} \label{optimal_at_utils}
    \begin{split} 
        \min_x & \; ||G x - \epsilon^*||_2 \\
        st & \; U(x) = x \rho \\
        st & \; x \preceq \gamma
    \end{split}
\end{equation}

\begin{algorithm} \label{alg_appro}
\caption{Sphere-Expo for approximating the Pareto-optimal solution space with hypersphere geodesic}
\begin{algorithmic}[1]
\Require 
    \Statex $\rho \in (0, 1)$: vector of relevance 
    \Statex $G \in \mathbb{R}^{n \times 2}, \: G_{ij} \in \{0, 1\}$: item masking group 
    \Statex $\varepsilon \in \mathbb{R}^{2}$: the target group fairness 
    \Statex $\gamma$: the position based exposure
    \Statex $n\_sample$: number of sample points in each geodesic curve
    \Statex $K$: number of marked points in logarithm of 2
\Ensure
    \Statex Pareto Set - $P$
    \Statex

\State $S \gets HyperSphere(dim=n-2)$
\State $start\_point \gets  \text{Solve (\ref{optimal_at_fairness}) for } \epsilon^*$
\State $end\_point \gets \gamma^{arg sort(\rho)}$
\State $q \gets [(start\_point, end\_point)]$
\State $P \gets []$
\While{$K > 0$}  
    \ForEach{$(s, e) \in q$}
        \State $geo \gets S.geodesic\_func(s \rightarrow e)$
        \State $mid\_point \gets geo.getMidPoint()$
        \State $mid\_util \gets utils(midpoint)$
        \State $fixed\_mid \gets  \text{Solve (\ref{optimal_at_utils}) for } mid\_util$ \qquad 
        \State $next\_q.append((s, fixed\_mid), (fixed\_mid, e))$
    \EndFor
    \State $q \gets next\_q$
    \State $K \gets K - 1$
\EndWhile

\ForEach{$(s, e) \in q$} \qquad $\triangleleft$ Sample point in geodesic
    \State $geo \gets S.geodesic\_func(s \rightarrow e)$
    \State $points \gets geo.sample(n\_sample)$
    \State $P.add(points)$
\EndFor
\end{algorithmic}
\end{algorithm}

\textit{Theoretical complexity. }
It is easily to see that the running time complexity of the Sphere-Expo algorithm mostly comes from solving (\ref{optimal_at_fairness}) for initiate point and solving (\ref{optimal_at_utils}) for marked points. Since the (\ref{optimal_at_utils}) is a constrained least squares problems, it could be solved in $\mathcal{O}(n^3\sqrt{n})$ \cite{doi:10.1287/moor.22.1.1} running time. The complexity for the LP problems (\ref{optimal_at_fairness}) might vary between solver methods but not larger than $\mathcal{O}(n^3\sqrt{n})$ \cite{ILLES2002170}. Therefore, the total complexity for the Sphere-Expo to approximate the whole Pareto-optimal space is $\mathcal{O}(K n^3\sqrt{n})$ with K is the number of marked points and n is the number of items.

\section{Experiments}
In order to solidify our empirical analysis of the Pareto-optimal solution space, we perform experiments on a small synthetic dataset. After that, we answer 2 question about the effectiveness and scalability of the Sphere-Expo approximation:
\begin{enumerate}
    \item How far is the approximation from complete Pareto optimality?
    \item How efficient is the approximation comparing to other baselines? 
\end{enumerate} 

\subsection{Experimental Settings}
\subsubsection{Datasets}
We perform experiments on two \emph{synthetic} datasets to answer the aforementioned questions and consolidate the efficiency of the Sphere-Expo approximation using two \emph{real-world} datasets.  

All \emph{synthetic} datasets are uniformly independently generated. For each number of items $n$, we sample the number of group $g$ and $n$ random relevance vectors whose elements are in the range $[0, 1]$. Each item is randomly assigned to a group using a binary matrix $G$. 

For the small \emph{synthetic} dataset $D_{s}$, the number of items $n$ is randomly sampled from the range of $n \in [8 \dots 20]$ with the number of groups of items being random. For the large \emph{synthetic} dataset $D_{l}$, the number of items $n$ is large $n \in {5 \dots 100}$ with the number of item-groups is random.

For \emph{real-world} datasets, we use \emph{TREC2020} Fair Ranking Dataset \cite{biega2021overview} and the \emph{MRLR} dataset \cite{DBLP:journals/corr/QinL13}. Documents in TREC are divided into different groups based on some criteria like released year, number of citations (divided in bins), etc. The documents in the MSLR dataset are divided into different groups based on the $132^{th}$ feature representing the quality score of a web page according to their documents. For TREC, we compute relevance probabilities ourselves, but for MSLR, we use the ground truth relevance scores graded from 0 (worst grade) to 4 (best grade), in order to check the influence of having discrete-valued relevance value (their relevance scores are normalized to the range [0, 1] by dividing them by 4). 

Queries with more than 100 documents are discarded due to the implementation problem: floating point error in the projection and \texttt{cvxpy} solver. We also eliminate uninteresting queries: queries having only one document, queries in which all relevance values are equal or queries having all the items belong to one group. The special cases when each item belongs to a different group: the item's individual fairness is discarded either. This leaves 183 queries in the TREC dataset and 861 queries in the MRLS dataset.   

\subsubsection {Baseline methods}
We evaluate our proposed methods and 3 baselines, in their ability to solve our MOO problem, both in terms of effectiveness and efficiency:
\begin{enumerate}
    \item A Quadratic Program (QP) that finds bi-stochastic matrices maximizing, for varying $\alpha$, the scalarized multi-objective function:
    \begin{equation} \label{QP}
    \begin{split}
        \max_B \alpha \rho^T B \gamma + (1-\alpha) ||B \gamma \times G - \epsilon^*||_2 \\
    st \; \mathbbm{1}B = \mathbbm{1}, \: B\mathbbm{1}^T = \mathbbm{1}
    \end{split}
    \end{equation}
    \item A static Controller (Ctrl) as proposed by \cite{MRFR}. This is a heuristic approach when each time a query is repeated, the disadvantaged items are compensated with a small amount of relevance scores so in the future, it will have higher chance to be at the top of the list. We customized the author's objectives to be suitable with our objectives' settings. 
    \item Our proposed path based on our theoretical profiling on the Expohedron facets: the P-Expo. 
    \item Our approximate Pareto front using the geodesic in the hypersphere: the Sphere-Expo. 
\end{enumerate}
While the QP baseline needs a Birkhoff a Neumann decomposition to generate the distribution of item orders before sampling, the P-Expo and Sphere-Expo could be combined with the Carathéodory algorithm proposed by \cite{Kletti_2022}. 

The package \texttt{cvxpy} \cite{cvxpy} is used for all linear and quadratic programming problems and the package \texttt{geomstats} \cite{geomstats} is used to generate the geodesic function on hypersphere. For all experiments, we use the DCG exposure vector for $\gamma$ whose elements at ranking $i^{th}$ is $\frac{1}{\log_2(i+2)}$ but any other PBMs should work.

\textit{Hyperparameters}
For the Ctrl, hyperparameter $K$ controls the balance between the efficiency (small $K$) and effectiveness (large K) is set to 1. By varying the gain $\lambda$ within $[0, 100]$, we expect the Ctrl could produce the Pareto front for each query. The gain $\alpha$ of the QP is also varied in the range $[0, 1]$ to find the optimal points. Results with the same gain are aggregated over queries to compute the final performances. 

While the P-Expo does not need any hyperparameter, the QP hyperparameter $\alpha$ in \ref{QP} varies between 0 and 1. Different hyperparameters $K$ in the Sphere-Expo representing the number of marked points will create different approximation curves for the Pareto front.  We will analyze how this hyperparameter affects our approximation in the next experiments.

Among our baselines, QP is guaranteed to produce Pareto-optimal points, provided that the associated QP solver has sufficient precision. The performances corresponding to the same hyperparameter $\alpha$ are aggregated over the different queries with the aggregation method. We will check whether QP solutions exist in our Expo-derived Pareto-front (P-Expo and Sphere-Expo) and associate the corresponding $\alpha$ to the latter. We are then able to aggregate the performances over the different queries, by considering the solutions associated with the same $\alpha$.

\subsection{Experimental results}
For each of the queries, we compute the full Pareto-front in the Expohedron using P-Expo. We also use Ctrl, QP and Sphere-Expo to find 20 Pareto optimal solutions.

\begin{figure}
    \centering
    \includegraphics[width=\columnwidth]{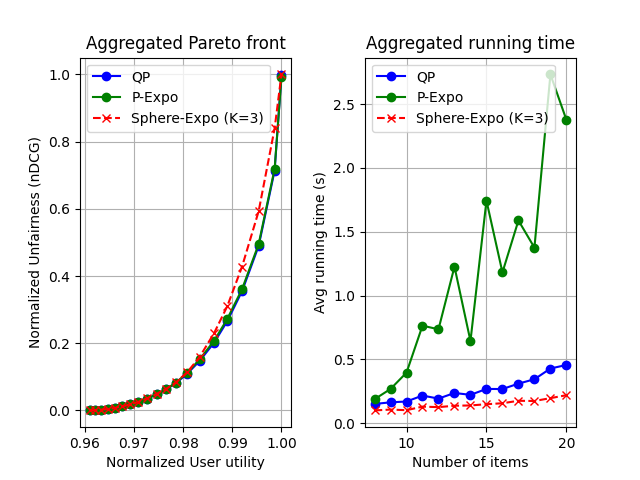}
    \caption{Aggregated Pareto fronts in small synthetic dataset $D_s$ showing the drawback of the P-Expo}
    \label{d_s}
\end{figure}

The experiment in the small synthetic dataset $D_s$ not only consolidates our corollary but also shows the drawback of the P-Expo. The aggregations of the Pareto front and their running time are reported in Figure [\ref{d_s}].

It appears without surprise that both QP and P-Expo perform identically. However, the total running of the P-Expo grows rapidly due to the fact that the number of segments (or facets it crosses) is unknown and incalculable. Therefore, it is not suitable to be scaled up and deployed in reality and is only used to verify our theoretical claims. On the other hand, the experimental results also show the potential of the approximation with Sphere-Expo. The red dash curve representing the Sphere-Expo with 7 marked is very close to the true Pareto-front. Therefore, in the next experiments, we will focus on analyzing the approximation using Sphere-Expo. 

\textit{How far is the approximation from complete Pareto optimality?} We perform experiments on the 2 real-world datasets TREC and MSLR to examine the performance of our proposed method in comparison with the baselines. 
\begin{figure}
    \centering
    \includegraphics[width=\linewidth]{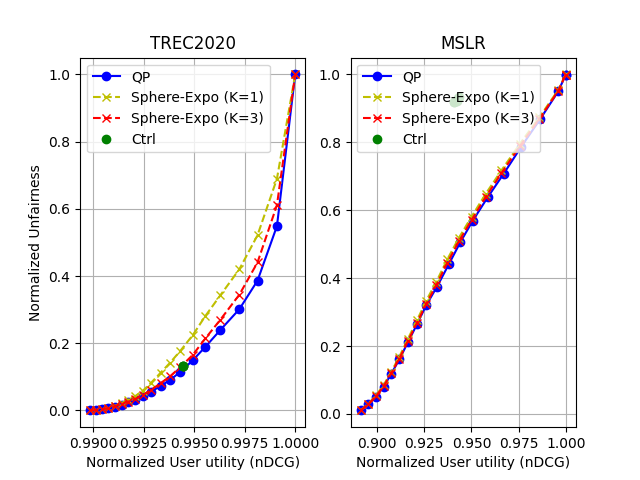}
    \caption{Aggregated Pareto fronts 2 real-world datasets. The more marked points, the closer the approximation to the true Pareto front.}
    \label{real_world}
\end{figure}

\begin{center} 
\begin{tabular}{ |c||c|c|} 
\hline
\textbf{Objectives} & \textbf{TREC} & \textbf{MSLR} \\
\hline
Normalized Utility (nDCG) &  0.995  & 0.94 \\
\hline
Normalized Unfairness & 0.132 & 0.93 \\
\hline
\end{tabular}
\captionof{table}{The normalized objectives in Ctrl baseline} \label{Ctrl_baseline}
\end{center}

The experiments in the two real-world datasets show that the fronts produced by our proposed method are moving forward to the true Pareto front when the number of marked points increases. Therefore, the Sphere-Expo could be used to approximate the true Pareto front. In the MSLR dataset, when the relevance scores are discrete values, the relevance vector tends to be orthogonal with some facets. Due to the asymmetric nature of the Expohedron, there will be more than one optimal solution, and the approximation is exactly matched with the true Pareto-front with one marked point. 

By varying the gain $\lambda$ in the Ctrl, we hope to find approximately Pareto-optimal solutions. However, after $T = 1000$, the aggregate results in both dataset always converge into one point. These points are denoted by green markers in the Figure [\ref{real_world}] and Table [\ref{Ctrl_baseline}]. In the MSLR dataset, the unfairness is extremely high in comparison with other methods. This might be due to the average number of items in the MSLR dataset is larger compared to the TREC dataset and the duplication in relevance score makes it hard to find an appropriate permutation in a limited time. 

\textit{How efficient is the approximation comparing to other baselines?}  

\begin{center} 
\begin{tabular}{ |c|c||c|c|} 
\hline
\textbf{Process} & \textbf{Algo} & \textbf{TREC} & \textbf{MSLR} \\
\hline
\multirow{4}{*}{Pareto front} & QP (20 points) & 1.23 & 4.45 \\
\cline{2-4}
& Sphere-Expo (K=1) & 0.28 & 0.56 \\ 
\cline{2-4}
& Sphere-Expo (K=3) & 0.47 & 0.99 \\ 
\hline
Decomposition & Carathéodory  & 0.07 & 0.54 \\
\cline{2-4}
& BvN & 2.25 & 29.56 \\
\hline
 \makecell{Full process \\ (target fairness)} & Ctrl & 0.47 & 1.39 \\
\hline
\end{tabular}
\captionof{table}{The average running time over queries to generate the Pareto front for $T = 1000$ in 2 real-world datasets} \label{real_world_running_time}
\end{center}

In Table [\ref{real_world_running_time}] we report the average time it takes to deliver $T = 1 000$ rankings for each query, for both datasets. Although the Ctrl algorithm does not need any initial cost, it needs to save and search history deliveries for the next ranking. Therefore, the Ctrl is faster for short time horizons T, and its running time will increase over time. Whereas for the Sphere-Expo and the QP methods, the computation of the Pareto-front and the decomposition need to be performed but once. Then, once a solution is found, it can be delivered very quickly using any sampling strategy. Therefore, for larger time horizons, the Sphere-Expo will be faster in total running time. 

\begin{figure}
    \centering
    \includegraphics[width=\linewidth]{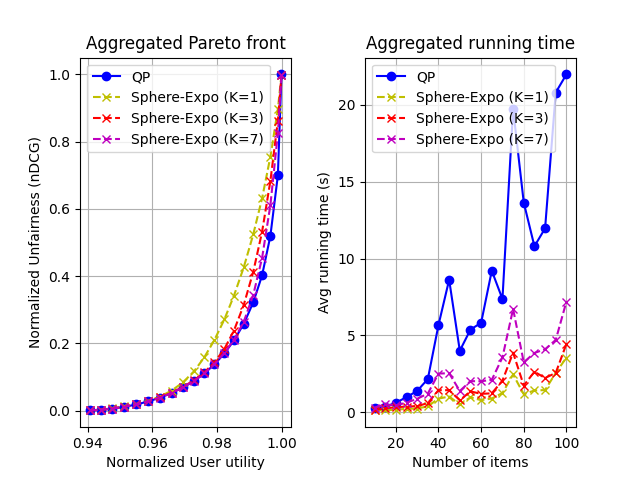}
    \caption{Aggregated Pareto fronts in large synthetic dataset $D_l$. Hyperparameter K denotes the number of marked points we sample in the approximation.}
    \label{d_l}
\end{figure}

We also investigate how the number of items affects the total running time for the QP baseline and the Sphere-Expo. In Figure [\ref{d_l}], we reported the accuracy and the total running time to generate 20 points in the Pareto front of the large synthetic dataset using QP and Sphere-Expo with different numbers of marked points. Obviously, the Sphere-Expo is asymptotic to the true Pareto-front when the number of marked points increases. The cost for this is the increase in the total running time. By varying the number of marked points sampling in the Sphere-Expo, we could control the trade-off between the accuracy and the time resources when approximating the Pareto front. Although our proposed method involves solving some quadratic programming for marked points, our problems have fewer variables and take fewer resources compared to the QP methods. Therefore, it is more controllable and could be adaptive to limited resource scenarios. 

\section{Conclusion}
Our novel geometrical algorithms allow us to approximate the true solutions space in Pareto-optimal utility-fairness among groups in repeated rankings' problem. Not only able to recover the Pareto frontier, our proposed methods are proved to be more efficient in two term: controllability and scalability. In the future work, we aim to expand our approximation method to other exposure models beyond PBMs such as Dynamic Bayesian Networks \cite{Chuklin2015ClickMF}.

\begin{acks}
This research was funded by Vingroup Innovation Foundation (VINIF) under project code VINIF.2022.DA00087
\end{acks}

\bibliographystyle{ACM-Reference-Format}
\bibliography{sample-base}


\begin{thebibliography}{43}


\ifx \showCODEN    \undefined \def \showCODEN     #1{\unskip}     \fi
\ifx \showDOI      \undefined \def \showDOI       #1{#1}\fi
\ifx \showISBNx    \undefined \def \showISBNx     #1{\unskip}     \fi
\ifx \showISBNxiii \undefined \def \showISBNxiii  #1{\unskip}     \fi
\ifx \showISSN     \undefined \def \showISSN      #1{\unskip}     \fi
\ifx \showLCCN     \undefined \def \showLCCN      #1{\unskip}     \fi
\ifx \shownote     \undefined \def \shownote      #1{#1}          \fi
\ifx \showarticletitle \undefined \def \showarticletitle #1{#1}   \fi
\ifx \showURL      \undefined \def \showURL       {\relax}        \fi
\providecommand\bibfield[2]{#2}
\providecommand\bibinfo[2]{#2}
\providecommand\natexlab[1]{#1}
\providecommand\showeprint[2][]{arXiv:#2}

\bibitem[Agrawal et~al\mbox{.}(2018)]%
        {cvxpy}
\bibfield{author}{\bibinfo{person}{Akshay Agrawal}, \bibinfo{person}{Robin Verschueren}, \bibinfo{person}{Steven Diamond}, {and} \bibinfo{person}{Stephen Boyd}.} \bibinfo{year}{2018}\natexlab{}.
\newblock \showarticletitle{A rewriting system for convex optimization problems}.
\newblock \bibinfo{journal}{\emph{Journal of Control and Decision}} \bibinfo{volume}{5}, \bibinfo{number}{1} (\bibinfo{year}{2018}), \bibinfo{pages}{42--60}.
\newblock


\bibitem[Arfken et~al\mbox{.}(1985)]%
        {ARFKEN200383}
\bibfield{author}{\bibinfo{person}{George~B. Arfken}, \bibinfo{person}{Hans~J. Weber}, {and} \bibinfo{person}{Frank~E. Harris}.} \bibinfo{year}{1985}\natexlab{}.
\newblock \showarticletitle{Chapter 2}.
\newblock In \bibinfo{booktitle}{\emph{Mathematical Methods for Physicists (Third Edition)} (\bibinfo{edition}{third edition} ed.)}, \bibfield{editor}{\bibinfo{person}{George~B. Arfken}, \bibinfo{person}{Hans~J. Weber}, {and} \bibinfo{person}{Frank~E. Harris}} (Eds.). \bibinfo{publisher}{Academic Press}, \bibinfo{address}{Orlando}, \bibinfo{pages}{102--111}.
\newblock
\showISBNx{978-0-12-384654-9}
\urldef\tempurl%
\url{https://doi.org/10.1016/B978-0-12-384654-9.00002-5}
\showDOI{\tempurl}


\bibitem[Aysha and Tarun(2022)]%
        {10.1007/978-981-16-2594-7_60}
\bibfield{author}{\bibinfo{person}{Saima Aysha} {and} \bibinfo{person}{Shrimali Tarun}.} \bibinfo{year}{2022}\natexlab{}.
\newblock \showarticletitle{A Pareto Dominance Approach to Multi-criteria Recommender System Using PSO Algorithm}. In \bibinfo{booktitle}{\emph{International Conference on Innovative Computing and Communications}}, \bibfield{editor}{\bibinfo{person}{Ashish Khanna}, \bibinfo{person}{Deepak Gupta}, \bibinfo{person}{Siddhartha Bhattacharyya}, \bibinfo{person}{Aboul~Ella Hassanien}, \bibinfo{person}{Sameer Anand}, {and} \bibinfo{person}{Ajay Jaiswal}} (Eds.). \bibinfo{publisher}{Springer Singapore}, \bibinfo{address}{Singapore}, \bibinfo{pages}{737--755}.
\newblock
\showISBNx{978-981-16-2594-7}


\bibitem[Biega et~al\mbox{.}(2021)]%
        {biega2021overview}
\bibfield{author}{\bibinfo{person}{Asia~J. Biega}, \bibinfo{person}{Fernando Diaz}, \bibinfo{person}{Michael~D. Ekstrand}, \bibinfo{person}{Sergey Feldman}, {and} \bibinfo{person}{Sebastian Kohlmeier}.} \bibinfo{year}{2021}\natexlab{}.
\newblock \bibinfo{title}{Overview of the TREC 2020 Fair Ranking Track}.
\newblock
\newblock
\showeprint[arxiv]{2108.05135}~[cs.IR]


\bibitem[Biega et~al\mbox{.}(2018)]%
        {Biega_2018}
\bibfield{author}{\bibinfo{person}{Asia~J. Biega}, \bibinfo{person}{Krishna~P. Gummadi}, {and} \bibinfo{person}{Gerhard Weikum}.} \bibinfo{year}{2018}\natexlab{}.
\newblock \showarticletitle{Equity of Attention}. In \bibinfo{booktitle}{\emph{The 41st International {ACM} {SIGIR} Conference on Research and Development in Information Retrieval}}. \bibinfo{publisher}{{ACM}}.
\newblock
\urldef\tempurl%
\url{https://doi.org/10.1145/3209978.3210063}
\showDOI{\tempurl}


\bibitem[Bishop(2013)]%
        {bishop2013riemannian}
\bibfield{author}{\bibinfo{person}{Richard~L. Bishop}.} \bibinfo{year}{2013}\natexlab{}.
\newblock \bibinfo{title}{Riemannian Geometry}.
\newblock
\newblock
\showeprint[arxiv]{1303.5390}~[math.DG]


\bibitem[Bowman(1972)]%
        {Permutahedron}
\bibfield{author}{\bibinfo{person}{V.~J. Bowman}.} \bibinfo{year}{1972}\natexlab{}.
\newblock \showarticletitle{Permutation Polyhedra}.
\newblock \bibinfo{journal}{\emph{SIAM J. Appl. Math.}} \bibinfo{volume}{22}, \bibinfo{number}{4} (\bibinfo{year}{1972}), \bibinfo{pages}{580--589}.
\newblock
\urldef\tempurl%
\url{https://doi.org/10.1137/0122054}
\showDOI{\tempurl}
\showeprint{https://doi.org/10.1137/0122054}


\bibitem[Carath{\'e}odory(1907)]%
        {Carathodory1907berDV}
\bibfield{author}{\bibinfo{person}{Constantin Carath{\'e}odory}.} \bibinfo{year}{1907}\natexlab{}.
\newblock \showarticletitle{{\"U}ber den Variabilit{\"a}tsbereich der Koeffizienten von Potenzreihen, die gegebene Werte nicht annehmen}.
\newblock \bibinfo{journal}{\emph{Math. Ann.}}  \bibinfo{volume}{64} (\bibinfo{year}{1907}), \bibinfo{pages}{95--115}.
\newblock
\urldef\tempurl%
\url{https://api.semanticscholar.org/CorpusID:116695038}
\showURL{%
\tempurl}


\bibitem[Chuklin and Markov(2015)]%
        {Chuklin2015ClickMF}
\bibfield{author}{\bibinfo{person}{Aleksandr Chuklin} {and} \bibinfo{person}{Ilya Markov}.} \bibinfo{year}{2015}\natexlab{}.
\newblock \showarticletitle{Click Models for Web Search Authors ’ version *}.
\newblock
\urldef\tempurl%
\url{https://api.semanticscholar.org/CorpusID:12961209}
\showURL{%
\tempurl}


\bibitem[Chuklin et~al\mbox{.}(2015)]%
        {book}
\bibfield{author}{\bibinfo{person}{Aleksandr Chuklin}, \bibinfo{person}{Ilya Markov}, {and} \bibinfo{person}{Maarten Rijke}.} \bibinfo{year}{2015}\natexlab{}.
\newblock \bibinfo{booktitle}{\emph{Click Models for Web Search}}. Vol.~\bibinfo{volume}{7}.
\newblock 1--115 pages.
\newblock
\urldef\tempurl%
\url{https://doi.org/10.2200/S00654ED1V01Y201507ICR043}
\showDOI{\tempurl}


\bibitem[Dahl(2010)]%
        {MajorizationTheory}
\bibfield{author}{\bibinfo{person}{Geir Dahl}.} \bibinfo{year}{2010}\natexlab{}.
\newblock \showarticletitle{Majorization permutahedra and (0,1)-matrices}.
\newblock \bibinfo{journal}{\emph{Linear Algebra Appl.}} \bibinfo{volume}{432}, \bibinfo{number}{12} (\bibinfo{year}{2010}), \bibinfo{pages}{3265--3271}.
\newblock
\showISSN{0024-3795}
\urldef\tempurl%
\url{https://doi.org/10.1016/j.laa.2010.01.024}
\showDOI{\tempurl}


\bibitem[Diaz et~al\mbox{.}(2020)]%
        {Diaz_2020}
\bibfield{author}{\bibinfo{person}{Fernando Diaz}, \bibinfo{person}{Bhaskar Mitra}, \bibinfo{person}{Michael~D. Ekstrand}, \bibinfo{person}{Asia~J. Biega}, {and} \bibinfo{person}{Ben Carterette}.} \bibinfo{year}{2020}\natexlab{}.
\newblock \showarticletitle{Evaluating Stochastic Rankings with Expected Exposure}. In \bibinfo{booktitle}{\emph{Proceedings of the 29th ACM International Conference on Information and Knowledge Management}} \emph{(\bibinfo{series}{CIKM ’20})}. \bibinfo{publisher}{ACM}.
\newblock
\urldef\tempurl%
\url{https://doi.org/10.1145/3340531.3411962}
\showDOI{\tempurl}


\bibitem[Do et~al\mbox{.}(2022)]%
        {Do_Corbett-Davies_Atif_Usunier_2022}
\bibfield{author}{\bibinfo{person}{Virginie Do}, \bibinfo{person}{Sam Corbett-Davies}, \bibinfo{person}{Jamal Atif}, {and} \bibinfo{person}{Nicolas Usunier}.} \bibinfo{year}{2022}\natexlab{}.
\newblock \showarticletitle{Online Certification of Preference-Based Fairness for Personalized Recommender Systems}.
\newblock \bibinfo{journal}{\emph{Proceedings of the AAAI Conference on Artificial Intelligence}} \bibinfo{volume}{36}, \bibinfo{number}{6} (\bibinfo{date}{Jun.} \bibinfo{year}{2022}), \bibinfo{pages}{6532--6540}.
\newblock
\urldef\tempurl%
\url{https://doi.org/10.1609/aaai.v36i6.20606}
\showDOI{\tempurl}


\bibitem[Dufossé and Uçar(2016)]%
        {Dufosse2016}
\bibfield{author}{\bibinfo{person}{Fanny Dufossé} {and} \bibinfo{person}{Bora Uçar}.} \bibinfo{year}{2016}\natexlab{}.
\newblock \showarticletitle{Notes on Birkhoff–von Neumann decomposition of doubly stochastic matrices}.
\newblock \bibinfo{journal}{\emph{Linear Algebra Appl.}}  \bibinfo{volume}{497} (\bibinfo{year}{2016}), \bibinfo{pages}{108--115}.
\newblock
\showISSN{0024-3795}
\urldef\tempurl%
\url{https://doi.org/10.1016/j.laa.2016.02.023}
\showDOI{\tempurl}


\bibitem[Fu et~al\mbox{.}(2021)]%
        {Popcorn}
\bibfield{author}{\bibinfo{person}{Zuohui Fu}, \bibinfo{person}{Yikun Xian}, \bibinfo{person}{Shijie Geng}, \bibinfo{person}{Gerard de Melo}, {and} \bibinfo{person}{Yongfeng Zhang}.} \bibinfo{year}{2021}\natexlab{}.
\newblock \showarticletitle{Popcorn: Human-in-the-Loop Popularity Debiasing in Conversational Recommender Systems}. In \bibinfo{booktitle}{\emph{Proceedings of the 30th ACM International Conference on Information and Knowledge Management}} (Virtual Event, Queensland, Australia) \emph{(\bibinfo{series}{CIKM '21})}. \bibinfo{publisher}{Association for Computing Machinery}, \bibinfo{address}{New York, NY, USA}, \bibinfo{pages}{494–503}.
\newblock
\showISBNx{9781450384469}
\urldef\tempurl%
\url{https://doi.org/10.1145/3459637.3482461}
\showDOI{\tempurl}


\bibitem[Ge et~al\mbox{.}(2021)]%
        {Ge_2021}
\bibfield{author}{\bibinfo{person}{Yingqiang Ge}, \bibinfo{person}{Shuchang Liu}, \bibinfo{person}{Ruoyuan Gao}, \bibinfo{person}{Yikun Xian}, \bibinfo{person}{Yunqi Li}, \bibinfo{person}{Xiangyu Zhao}, \bibinfo{person}{Changhua Pei}, \bibinfo{person}{Fei Sun}, \bibinfo{person}{Junfeng Ge}, \bibinfo{person}{Wenwu Ou}, {and} \bibinfo{person}{Yongfeng Zhang}.} \bibinfo{year}{2021}\natexlab{}.
\newblock \showarticletitle{Towards Long-term Fairness in Recommendation}. In \bibinfo{booktitle}{\emph{Proceedings of the 14th ACM International Conference on Web Search and Data Mining}} \emph{(\bibinfo{series}{WSDM ’21})}. \bibinfo{publisher}{ACM}.
\newblock
\urldef\tempurl%
\url{https://doi.org/10.1145/3437963.3441824}
\showDOI{\tempurl}


\bibitem[Ge et~al\mbox{.}(2022)]%
        {Ge_2022}
\bibfield{author}{\bibinfo{person}{Yingqiang Ge}, \bibinfo{person}{Xiaoting Zhao}, \bibinfo{person}{Lucia Yu}, \bibinfo{person}{Saurabh Paul}, \bibinfo{person}{Diane Hu}, \bibinfo{person}{Chu-Cheng Hsieh}, {and} \bibinfo{person}{Yongfeng Zhang}.} \bibinfo{year}{2022}\natexlab{}.
\newblock \showarticletitle{Toward Pareto Efficient Fairness-Utility Trade-off in Recommendation through Reinforcement Learning}. In \bibinfo{booktitle}{\emph{Proceedings of the Fifteenth ACM International Conference on Web Search and Data Mining}} \emph{(\bibinfo{series}{WSDM ’22})}. \bibinfo{publisher}{ACM}.
\newblock
\urldef\tempurl%
\url{https://doi.org/10.1145/3488560.3498487}
\showDOI{\tempurl}


\bibitem[Geyik et~al\mbox{.}(2019)]%
        {Geyik_2019}
\bibfield{author}{\bibinfo{person}{Sahin~Cem Geyik}, \bibinfo{person}{Stuart Ambler}, {and} \bibinfo{person}{Krishnaram Kenthapadi}.} \bibinfo{year}{2019}\natexlab{}.
\newblock \showarticletitle{Fairness-Aware Ranking in Search Recommendation Systems with Application to {LinkedIn} Talent Search}. In \bibinfo{booktitle}{\emph{Proceedings of the 25th {ACM} {SIGKDD} International Conference on Knowledge Discovery and Data Mining}}. \bibinfo{publisher}{ACM}.
\newblock
\urldef\tempurl%
\url{https://doi.org/10.1145/3292500.3330691}
\showDOI{\tempurl}


\bibitem[Gotoh(1994)]%
        {Gotoh94}
\bibfield{author}{\bibinfo{person}{Tohru Gotoh}.} \bibinfo{year}{1994}\natexlab{}.
\newblock \showarticletitle{GEODESIC HYPERSPHERES IN COMPLEX PROJECTIVE SPACE}.
\newblock \bibinfo{journal}{\emph{Tsukuba Journal of Mathematics}} \bibinfo{volume}{18}, \bibinfo{number}{1} (\bibinfo{year}{1994}), \bibinfo{pages}{207--215}.
\newblock
\showISSN{03874982}
\urldef\tempurl%
\url{http://www.jstor.org/stable/43685888}
\showURL{%
\tempurl}


\bibitem[Hilton(1964)]%
        {topology}
\bibfield{author}{\bibinfo{person}{P.~J. Hilton}.} \bibinfo{year}{1964}\natexlab{}.
\newblock \showarticletitle{Topology. By John G. Hocking and Gail S. Young. Pp. 382. 74s. 1961. (Addison-Wesley)}.
\newblock \bibinfo{journal}{\emph{The Mathematical Gazette}} \bibinfo{volume}{48}, \bibinfo{number}{363} (\bibinfo{year}{1964}), \bibinfo{pages}{122–123}.
\newblock
\urldef\tempurl%
\url{https://doi.org/10.2307/3614368}
\showDOI{\tempurl}


\bibitem[Horen(1985)]%
        {LP}
\bibfield{author}{\bibinfo{person}{Jeff Horen}.} \bibinfo{year}{1985}\natexlab{}.
\newblock \showarticletitle{Linear programming, by Katta G. Murty, John Wiley and Sons, New York, 1983, 482 pp.}
\newblock \bibinfo{journal}{\emph{Networks}} \bibinfo{volume}{15}, \bibinfo{number}{2} (\bibinfo{year}{1985}), \bibinfo{pages}{273--274}.
\newblock
\urldef\tempurl%
\url{http://dblp.uni-trier.de/db/journals/networks/networks15.html#Horen85}
\showURL{%
\tempurl}


\bibitem[Illés and Terlaky(2002)]%
        {ILLES2002170}
\bibfield{author}{\bibinfo{person}{Tibor Illés} {and} \bibinfo{person}{Tamás Terlaky}.} \bibinfo{year}{2002}\natexlab{}.
\newblock \showarticletitle{Pivot versus interior point methods: Pros and cons}.
\newblock \bibinfo{journal}{\emph{European Journal of Operational Research}} \bibinfo{volume}{140}, \bibinfo{number}{2} (\bibinfo{year}{2002}), \bibinfo{pages}{170--190}.
\newblock
\showISSN{0377-2217}
\urldef\tempurl%
\url{https://doi.org/10.1016/S0377-2217(02)00061-9}
\showDOI{\tempurl}


\bibitem[J\"{a}rvelin and Kek\"{a}l\"{a}inen(2002)]%
        {DCG}
\bibfield{author}{\bibinfo{person}{Kalervo J\"{a}rvelin} {and} \bibinfo{person}{Jaana Kek\"{a}l\"{a}inen}.} \bibinfo{year}{2002}\natexlab{}.
\newblock \showarticletitle{Cumulated Gain-Based Evaluation of IR Techniques}.
\newblock \bibinfo{journal}{\emph{ACM Trans. Inf. Syst.}} \bibinfo{volume}{20}, \bibinfo{number}{4} (\bibinfo{date}{oct} \bibinfo{year}{2002}), \bibinfo{pages}{422–446}.
\newblock
\showISSN{1046-8188}
\urldef\tempurl%
\url{https://doi.org/10.1145/582415.582418}
\showDOI{\tempurl}


\bibitem[Jin et~al\mbox{.}(2023)]%
        {Jin2023ParetobasedMR}
\bibfield{author}{\bibinfo{person}{Jipeng Jin}, \bibinfo{person}{Zhaoxiang Zhang}, \bibinfo{person}{Zhiheng Li}, \bibinfo{person}{Xiaofeng Gao}, \bibinfo{person}{Xiongwei Yang}, \bibinfo{person}{Lei Xiao}, {and} \bibinfo{person}{Jie Jiang}.} \bibinfo{year}{2023}\natexlab{}.
\newblock \showarticletitle{Pareto-based Multi-Objective Recommender System with Forgetting Curve}.
\newblock \bibinfo{journal}{\emph{ArXiv}}  \bibinfo{volume}{abs/2312.16868} (\bibinfo{year}{2023}).
\newblock
\urldef\tempurl%
\url{https://api.semanticscholar.org/CorpusID:266573337}
\showURL{%
\tempurl}


\bibitem[Kletti et~al\mbox{.}(2022)]%
        {Kletti_2022}
\bibfield{author}{\bibinfo{person}{Till Kletti}, \bibinfo{person}{Jean-Michel Renders}, {and} \bibinfo{person}{Patrick Loiseau}.} \bibinfo{year}{2022}\natexlab{}.
\newblock \showarticletitle{Introducing the Expohedron for Efficient Pareto-optimal Fairness-Utility Amortizations in Repeated Rankings}. In \bibinfo{booktitle}{\emph{Proceedings of the Fifteenth {ACM} International Conference on Web Search and Data Mining}}. \bibinfo{publisher}{{ACM}}.
\newblock
\urldef\tempurl%
\url{https://doi.org/10.1145/3488560.3498490}
\showDOI{\tempurl}


\bibitem[Marshall et~al\mbox{.}(1980)]%
        {majorizarion}
\bibfield{author}{\bibinfo{person}{Albert~W. Marshall}, \bibinfo{person}{Ingram Olkin}, {and} \bibinfo{person}{Barry~C. Arnold}.} \bibinfo{year}{1980}\natexlab{}.
\newblock \showarticletitle{Inequalities: Theory of Majorization and Its Applications}.
\newblock
\urldef\tempurl%
\url{https://api.semanticscholar.org/CorpusID:37387169}
\showURL{%
\tempurl}


\bibitem[Moffat and Zobel(2008)]%
        {RPB}
\bibfield{author}{\bibinfo{person}{Alistair Moffat} {and} \bibinfo{person}{Justin Zobel}.} \bibinfo{year}{2008}\natexlab{}.
\newblock \showarticletitle{Rank-Biased Precision for Measurement of Retrieval Effectiveness}.
\newblock \bibinfo{journal}{\emph{ACM Trans. Inf. Syst.}} \bibinfo{volume}{27}, \bibinfo{number}{1}, Article \bibinfo{articleno}{2} (\bibinfo{date}{dec} \bibinfo{year}{2008}), \bibinfo{numpages}{27}~pages.
\newblock
\showISSN{1046-8188}
\urldef\tempurl%
\url{https://doi.org/10.1145/1416950.1416952}
\showDOI{\tempurl}


\bibitem[Naghiaei et~al\mbox{.}(2022)]%
        {Naghiaei2022TheUO}
\bibfield{author}{\bibinfo{person}{Mohammadmehdi Naghiaei}, \bibinfo{person}{Hossein~A. Rahmani}, {and} \bibinfo{person}{Mahdi Dehghan}.} \bibinfo{year}{2022}\natexlab{}.
\newblock \showarticletitle{The Unfairness of Popularity Bias in Book Recommendation}. In \bibinfo{booktitle}{\emph{International Workshop on Algorithmic Bias in Search and Recommendation}}.
\newblock
\urldef\tempurl%
\url{https://api.semanticscholar.org/CorpusID:247158125}
\showURL{%
\tempurl}


\bibitem[Nesterov and Todd(1997)]%
        {doi:10.1287/moor.22.1.1}
\bibfield{author}{\bibinfo{person}{Yu.~E. Nesterov} {and} \bibinfo{person}{M.~J. Todd}.} \bibinfo{year}{1997}\natexlab{}.
\newblock \showarticletitle{Self-Scaled Barriers and Interior-Point Methods for Convex Programming}.
\newblock \bibinfo{journal}{\emph{Mathematics of Operations Research}} \bibinfo{volume}{22}, \bibinfo{number}{1} (\bibinfo{year}{1997}), \bibinfo{pages}{1--42}.
\newblock
\urldef\tempurl%
\url{https://doi.org/10.1287/moor.22.1.1}
\showDOI{\tempurl}
\showeprint{https://doi.org/10.1287/moor.22.1.1}


\bibitem[Paparella(2022)]%
        {Paparella2022PursuingOT}
\bibfield{author}{\bibinfo{person}{Vincenzo Paparella}.} \bibinfo{year}{2022}\natexlab{}.
\newblock \showarticletitle{Pursuing Optimal Trade-Off Solutions in Multi-Objective Recommender Systems}.
\newblock \bibinfo{journal}{\emph{Proceedings of the 16th ACM Conference on Recommender Systems}} (\bibinfo{year}{2022}).
\newblock
\urldef\tempurl%
\url{https://api.semanticscholar.org/CorpusID:252216518}
\showURL{%
\tempurl}


\bibitem[Pareto(2014)]%
        {ParetoIntro}
\bibfield{author}{\bibinfo{person}{Vilfredo Pareto}.} \bibinfo{year}{2014}\natexlab{}.
\newblock \bibinfo{booktitle}{\emph{{Manual of Political Economy: A Critical and Variorum Edition}}}.
\newblock Number 9780199607952 in \bibinfo{series}{OUP Catalogue}. \bibinfo{publisher}{Oxford University Press}.
\newblock
\showISBNx{ARRAY(0x59c17ac0)}
\urldef\tempurl%
\url{https://ideas.repec.org/b/oxp/obooks/9780199607952.html}
\showURL{%
\tempurl}


\bibitem[Qin and Liu(2013)]%
        {DBLP:journals/corr/QinL13}
\bibfield{author}{\bibinfo{person}{Tao Qin} {and} \bibinfo{person}{Tie{-}Yan Liu}.} \bibinfo{year}{2013}\natexlab{}.
\newblock \showarticletitle{Introducing {LETOR} 4.0 Datasets}.
\newblock \bibinfo{journal}{\emph{CoRR}}  \bibinfo{volume}{abs/1306.2597} (\bibinfo{year}{2013}).
\newblock
\urldef\tempurl%
\url{http://arxiv.org/abs/1306.2597}
\showURL{%
\tempurl}


\bibitem[Salimi et~al\mbox{.}(2019)]%
        {salimi2019capuchin}
\bibfield{author}{\bibinfo{person}{Babak Salimi}, \bibinfo{person}{Luke Rodriguez}, \bibinfo{person}{Bill Howe}, {and} \bibinfo{person}{Dan Suciu}.} \bibinfo{year}{2019}\natexlab{}.
\newblock \bibinfo{title}{Capuchin: Causal Database Repair for Algorithmic Fairness}.
\newblock
\newblock
\showeprint[arxiv]{1902.08283}~[cs.DB]


\bibitem[Seymen et~al\mbox{.}(2021)]%
        {SeymenAM21a}
\bibfield{author}{\bibinfo{person}{Sinan Seymen}, \bibinfo{person}{Himan Abdollahpouri}, {and} \bibinfo{person}{Edward~C. Malthouse}.} \bibinfo{year}{2021}\natexlab{}.
\newblock \showarticletitle{A Unified Optimization Toolbox for Solving Popularity Bias, Fairness, and Diversity in Recommender Systems}. In \bibinfo{booktitle}{\emph{Proceedings of the 1st Workshop on Multi-Objective Recommender Systems (MORS 2021) co-located with 15th ACM Conference on Recommender Systems (RecSys 2021), Amsterdam, The Netherlands, September 25, 2021}} \emph{(\bibinfo{series}{CEUR Workshop Proceedings}, Vol.~\bibinfo{volume}{2959})}, \bibfield{editor}{\bibinfo{person}{Himan Abdollahpouri}, \bibinfo{person}{Mehdi Elahi}, \bibinfo{person}{Masoud Mansoury}, \bibinfo{person}{Shaghayegh Sahebi}, \bibinfo{person}{Zahra Nazari}, \bibinfo{person}{Allison Chaney}, {and} \bibinfo{person}{Babak Loni}} (Eds.). \bibinfo{publisher}{CEUR-WS.org}.
\newblock
\urldef\tempurl%
\url{http://ceur-ws.org/Vol-2959/paper5.pdf}
\showURL{%
\tempurl}


\bibitem[Singh and Joachims(2018)]%
        {Singh_2018}
\bibfield{author}{\bibinfo{person}{Ashudeep Singh} {and} \bibinfo{person}{Thorsten Joachims}.} \bibinfo{year}{2018}\natexlab{}.
\newblock \showarticletitle{Fairness of Exposure in Rankings}. In \bibinfo{booktitle}{\emph{Proceedings of the 24th {ACM} {SIGKDD} International Conference on Knowledge Discovery and Data Mining}}. \bibinfo{publisher}{{ACM}}.
\newblock
\urldef\tempurl%
\url{https://doi.org/10.1145/3219819.3220088}
\showDOI{\tempurl}


\bibitem[Snyder(2012)]%
        {Snyder2012MapPA}
\bibfield{author}{\bibinfo{person}{John~P. Snyder}.} \bibinfo{year}{2012}\natexlab{}.
\newblock \showarticletitle{Map Projections: A Working Manual}.
\newblock
\urldef\tempurl%
\url{https://api.semanticscholar.org/CorpusID:60447053}
\showURL{%
\tempurl}


\bibitem[Su et~al\mbox{.}(2021)]%
        {su2021optimizing}
\bibfield{author}{\bibinfo{person}{Yi Su}, \bibinfo{person}{Magd Bayoumi}, {and} \bibinfo{person}{Thorsten Joachims}.} \bibinfo{year}{2021}\natexlab{}.
\newblock \bibinfo{title}{Optimizing Rankings for Recommendation in Matching Markets}.
\newblock
\newblock
\showeprint[arxiv]{2106.01941}~[cs.IR]


\bibitem[team({[n.\,d.]})]%
        {geomstats}
\bibfield{author}{\bibinfo{person}{Geomstats team}.} \bibinfo{year}{[n.\,d.]}\natexlab{}.
\newblock \bibinfo{booktitle}{\emph{https://geomstats.github.io/}}.
\newblock
\urldef\tempurl%
\url{https://geomstats.github.io/}
\showURL{%
\tempurl}


\bibitem[Thonet and Renders(2020)]%
        {MRFR}
\bibfield{author}{\bibinfo{person}{Thibaut Thonet} {and} \bibinfo{person}{Jean-Michel Renders}.} \bibinfo{year}{2020}\natexlab{}.
\newblock \showarticletitle{Multi-Grouping Robust Fair Ranking}. In \bibinfo{booktitle}{\emph{Proceedings of the 43rd International ACM SIGIR Conference on Research and Development in Information Retrieval}} (Virtual Event, China) \emph{(\bibinfo{series}{SIGIR '20})}. \bibinfo{publisher}{Association for Computing Machinery}, \bibinfo{address}{New York, NY, USA}, \bibinfo{pages}{2077–2080}.
\newblock
\showISBNx{9781450380164}
\urldef\tempurl%
\url{https://doi.org/10.1145/3397271.3401292}
\showDOI{\tempurl}


\bibitem[Wang and Joachims(2021)]%
        {Wang_2021}
\bibfield{author}{\bibinfo{person}{Lequn Wang} {and} \bibinfo{person}{Thorsten Joachims}.} \bibinfo{year}{2021}\natexlab{}.
\newblock \showarticletitle{User Fairness, Item Fairness, and Diversity for Rankings in Two-Sided Markets}. In \bibinfo{booktitle}{\emph{Proceedings of the 2021 ACM SIGIR International Conference on Theory of Information Retrieval}} \emph{(\bibinfo{series}{ICTIR ’21})}. \bibinfo{publisher}{ACM}.
\newblock
\urldef\tempurl%
\url{https://doi.org/10.1145/3471158.3472260}
\showDOI{\tempurl}


\bibitem[Wang et~al\mbox{.}(2020)]%
        {Wang2020DCNVI}
\bibfield{author}{\bibinfo{person}{Ruoxi Wang}, \bibinfo{person}{Rakesh Shivanna}, \bibinfo{person}{Derek~Zhiyuan Cheng}, \bibinfo{person}{Sagar Jain}, \bibinfo{person}{Dong Lin}, \bibinfo{person}{Lichan Hong}, {and} \bibinfo{person}{Ed~H. Chi}.} \bibinfo{year}{2020}\natexlab{}.
\newblock \showarticletitle{DCN V2: Improved Deep and Cross Network and Practical Lessons for Web-scale Learning to Rank Systems}.
\newblock \bibinfo{journal}{\emph{Proceedings of the Web Conference 2021}} (\bibinfo{year}{2020}).
\newblock
\urldef\tempurl%
\url{https://api.semanticscholar.org/CorpusID:224845398}
\showURL{%
\tempurl}


\bibitem[Zafar et~al\mbox{.}(2019)]%
        {10.5555/3322706.3362016}
\bibfield{author}{\bibinfo{person}{Muhammad~Bilal Zafar}, \bibinfo{person}{Isabel Valera}, \bibinfo{person}{Manuel Gomez-Rodriguez}, {and} \bibinfo{person}{Krishna~P. Gummadi}.} \bibinfo{year}{2019}\natexlab{}.
\newblock \showarticletitle{Fairness Constraints: A Flexible Approach for Fair Classification}.
\newblock \bibinfo{journal}{\emph{J. Mach. Learn. Res.}} \bibinfo{volume}{20}, \bibinfo{number}{1} (\bibinfo{date}{jan} \bibinfo{year}{2019}), \bibinfo{pages}{2737–2778}.
\newblock
\showISSN{1532-4435}


\bibitem[Zerveas(2022)]%
        {Zerveas2022MitigatingBI}
\bibfield{author}{\bibinfo{person}{George Zerveas}.} \bibinfo{year}{2022}\natexlab{}.
\newblock \showarticletitle{Mitigating Bias in Search Results Through Contextual Document Reranking and Neutrality Regularization}.
\newblock \bibinfo{journal}{\emph{Proceedings of the 45th International ACM SIGIR Conference on Research and Development in Information Retrieval}} (\bibinfo{year}{2022}).
\newblock
\urldef\tempurl%
\url{https://api.semanticscholar.org/CorpusID:248837769}
\showURL{%
\tempurl}


\end{thebibliography}

\end{document}